\documentstyle[12pt]{article}
\newcommand\bb{\begin{equation}}
\newcommand\ee{\end{equation}}
\def\pd{\partial}
\newcommand\bba{\begin{eqnarray}}
\newcommand\eea{\end{eqnarray}}

\begin{document}

\rightline
\bigskip
\centerline{\bf A Model of Quantum Electrodynamics with Higher
Derivatives}
\bigskip\bigskip
\centerline{\it D.Ts. Stoyanov\footnote{This work is supported in
part by Bulgarian National Foundation for Scientific Research
Ph-401.}}
\medskip
\centerline{Institute for Nuclear Research and Nuclear Energy}
\centerline{Boul.Tsarigradsko ch. 72}
\centerline{1784 Sofia Bulgaria\footnote{e-mail
dstoyan@bgearn.acad.bg}}
\bigskip\bigskip\bigskip\bigskip
\centerline{\bf Abstract}

A new version of application Pauli-Villars regularized Green
functions in the quantum field theory using
higher derivatives is proposed. In this version the regularizing
mass $M$ is large but finite.
Our approach is demonstrated and discussed on the example of QED.
It is shown that in our case there are no ultraviolet divergences
and - on the example of the selfenergy spinor
Feynman diagram - no infrared ones.

\bigskip

\bigskip\bigskip

\newpage

\section{Introduction}

One of the unwritten rules of the standard quantum field theory is
the limited order of the derivatives in the field equations. As is
well known this order is two for the field with integer spin and -
one  for  half  spin.  This  rule  appears,  at  first,  with  the
Klein-Gordon and Dirac equations, which stand in the origin of
many
field models. We can suppose that the analogy with  the  classical
mechanics is one of the main reason for this situation  -  let  us
recall that only first order time derivatives enter the  action  of
the usual mechanical systems. However, mechanical actions  with
higher time derivatives were considered sometimes by different
authors \cite {r1}. In this case they use  the  special  canonical
formalism first developed by Ostrogradski \cite {r2}

Recently models with higher derivatives appeared also in
quantum field theory, especially in the two dimensional one (see
for example \cite{r3} where the reader can find other references
on this subject). However, we are going to concentrate our
attention to the use of higher derivatives in the 3+1 dimensional
quantum field theory. In this direction we can point out
works in which the authors study several mathematical properties
of the free field equations with higher derivatives. These
equations are characterized by a polynomial in the d'Alambertian
operator \cite{r4}. It is interesting to mention the work
\cite{r5},  where the higher derivative kinetic term was
introduced in the Nambu-Jona-Lasinio model.

In the last few years appeared several works in which the authors
study the so called "differential regularization" first proposed by
Freedman, Johnson and Latorre \cite{r6}. Though the derivatives
are used there only for the regularization of the loop diagram in
the coordinate representation, we may suppose that these formal
rules maybe follow from the appropriate quantum field theory
with higher derivatives.

The main difficulty arising when we use higher derivatives in the
quantum field theory is the appearance of the indefinite metric
in the state space and all physical consequences from this. That
is why quantum field theory with higher derivatives in the
free part of the corresponding Lagrangean can not exist, if we
want every vector from the state space to have a physical
meaning (here we have in mind the spaces of in- and out- states).
However, there exists another approach in which the physical
states
form the appropriate subspace with definite metric in the frame
of the whole indefinite state space. Such an approach was
sketched
by Hawking \cite{r7}. In this case the physical theory appears
as immersed in a wider theory, analogously to the situation in
the quantum electrodynamics in the Lorentz gauge.

Different authors adduce different arguments for the
application of higher derivatives depending on the concrete problems
they pay attention to. The presence of higher derivatives in the
kinetic part of the field Lagrangean leads to free
propagation function in which the power of the momentum $p$ is
less than -2. This fact means that the divergences of the Feynman
diagrams will become smaller. For example if the propagator of
the fermions in the usual spinor QED has the behaviour $p^{-3}$,
when $p\to\infty$, then all well known divergent diagrams become
finite. This fact is the main argument for us to consider higher
derivatives. The aim of the present paper is to give an
alternative formulation of the spinor QED in which the
fermion field obeys a third order differential equation.

\section{The Model}

We are going to consider a model in which the main fields are the
spinor field $\varphi(x)$ and vector one $A_\mu(x)$.
Besides we have
an auxiliary spinor and scalar fields denoted by $\psi(x)$ and
$\Phi(x)$ respectively. To write down the action $S$ of our
model, let us introduce the following notation
\bb
D_\mu=\pd_\mu-ieA_\mu\ \ \ \ \ \nabla_\mu=\pd_\mu-ie\pd_\mu\Phi
\label{eq1}
\ee
where $e$ is the dimensionless electric charge.
(From  here on the Greek indices such as
($\mu,\nu,\lambda$) run from 0 to 3 -  as vector indices
- those such as ($\alpha, \beta, \gamma$) run from 1
to 4 - as spinor indices). Then
\bb
  S=\int[L_{sp}(\phi)+L_v(A)+L^T(\phi)]d^4x
\label{eq2}
\ee
where the parts $L_{sp}$ and $L_v$ of the Lagrangean have the
following form:
\bba
L_{sp}(\varphi)&=&{1\over\sqrt{M^2-m^2}} (\nabla^\star_\mu{\bar\varphi}
\nabla^\mu\psi + \nabla^\star_\mu{\bar\psi}\nabla^\mu\varphi) +
{\bar\varphi}(i/2\gamma^\mu D^{\!\!\!\!\!\! ^\leftrightarrow}_\mu +
 m) \varphi+ \nonumber \\ &&+
{\bar\psi}(i/2\gamma^\mu \nabla^{\!\!\!\!\!\! ^\leftrightarrow}_\mu
 - m)\psi -
{m^2\over\sqrt{M^2-m^2}}({\bar\psi}\varphi+{\bar\varphi}\psi)
\label{eq3}\\
L_v(A)&=&-1/2\pd^\mu(A_\nu-\pd_\nu\Phi)\pd_\mu(A^\nu-\pd^\nu\Phi)
\label{eq4}
\eea
$m$ and $M$ are mass parameters.
Here with $\nabla^\star_\mu$ and $\bar\varphi$ we have denoted
the complex and Dirac conjugation respectively. Moreover we have
denoted:
$$ u{D^{\!\!\!\!\!\!^\leftrightarrow}_\mu}v=u{D_\mu}v-{D_\mu}u.v$$
$$ u{\nabla^{\!\!\!\!\!\!^\leftrightarrow}_\mu}v=u{\nabla_\mu}v
-{\nabla_\mu}u.v $$
$L^T(\Phi)$ is the free kinetic part of our Lagrangean for the
scalar field $\Phi$. As we can see below the concrete form of
this term does not matter for our model.

For the construction of the Lagrangean (\ref{eq3}) we have used, in
general, a known manipulation for the exclusion of the highest
derivative ( in our case - third derivative), with the help of
the auxiliary field $\psi(x)$. It is easy to verify, that if we
replace the field $\psi(x)$ by new a auxiliary field $\chi(x)$ with
the relation
\bb
\psi(x)=\chi(x)-{1\over\sqrt{M^2-m^2}}(i{\gamma^\mu}\nabla_\mu+m)
\varphi(x)
\label{eq5}
\ee
then we can obtain the following new form of the spinor part of
the Lagrangean
\bba
L_{sp}(\varphi)&=&-{1\over{M^2-m^2}}{\nabla^\star_\mu}\bar\varphi
(i/2\gamma^\mu \nabla^{\!\!\!\!\!\! ^\leftrightarrow}_\mu
 + m){\nabla^\mu}\varphi \nonumber \\
 &+&
 {m^2\over{M^2-m^2}}\bar\varphi
(i/2\gamma^\mu \nabla^{\!\!\!\!\!\! ^\leftrightarrow}_\mu
 + m)\varphi  \nonumber \\
 &+&
 \bar\varphi(i/2\gamma^\mu D^{\!\!\!\!\!\! ^\leftrightarrow}_\mu +
 m)\varphi+\bar\chi
 (i/2\gamma^\mu \nabla^{\!\!\!\!\!\! ^\leftrightarrow}_\mu
 - m)\chi
\label{eq6}
\eea
without the change of the action (\ref{eq2}). We see that the free
field $\chi$ is not connected with the rest of the fields and has
no significance
for our model. In spite of the equivalence between the spinor
part (\ref{eq3}) of the Lagrangean and $L_{sp}(\varphi)$ from
(\ref{eq6})
we give our preferences to the one from eq. (\ref{eq3}), because
it is
more suitable for the applications of the canonical quantum
formalism.

The Lagrangean parts $L_{sp}(\varphi)$ and $L_{v}(A)$ are gauge
invariant under the following local transformations:
\bba
&\varphi(x)\rightarrow\exp{ie\eta(x)}.\varphi(x);
&\psi(x)\rightarrow\exp{ie\eta(x)}.\psi(x)  \nonumber \\
&{A_\mu}(x)\rightarrow {A_\mu}(x)+{\pd_\mu}\eta(x);
&\Phi(x)\rightarrow \Phi(x)+\eta(x)
\label{eq7}
\eea
where $\eta(x)$ is the corresponding local parameter.

Let us try then to write down $L^{T}(\Phi)$
in the gauge invariant
form. In this case it has to depend on ${\pd_\mu}\Phi-A_\mu$,
because this is the only gauge invariant combination in  the
presence of the field $\Phi$:
\bb
{L^T}(\Phi)=W({\pd_\mu}\Phi-A_\mu)
\label{eq8}
\ee
In this case the action (\ref{eq2}) becomes gauge invariant.
However, with a simple redefinition of the fields $\psi,\varphi$
and $A_\mu$ with the help of the gauge transformations
(\ref{eq7}), the field $\Phi$ will disappear from the action. This
means that the free Lagrangean ${L^T}(\Phi)$ can not be gauge
invariant for arbitrary $\eta(x)$. Taking into account that
the conformal dimension of $\Phi(x)$ is zero, it is natural to
suppose that the free part of its equation must be ${\Box^2}\Phi$
($\Box$ is d'Alambertian operator). Then the gauge
invariance is of the type (\ref{eq7}) but
with $\eta(x)$ satisfying the equation
\bb
{\Box^2}\eta(x) \label{eq9}=0
\ee
As we shall see below this restriction is in accordance with the
usually used gauge fixing rules in QED.

Now we can give the explicit form of ${L^T}(\Phi)$:
\bb
{L^T}(\Phi)={\pd^\mu}\Phi{\pd_\mu}U+{1\over{2}}U^2
\label{eq10}
\ee
$U(x)$ is an auxiliary field with the help of which ${L^T}(\Phi)$
depends on first derivatives only. The gauge transformation of
$U(x)$ has the form
$$U(x)\rightarrow{U(x)+\Box\eta(x)}$$
Then we can obtain the field equations. Varying the action (\ref{eq2})
we have the primary equations:
\bb
\sqrt{M^2-m^2}(i{\gamma^\mu}D_\mu+m)\varphi=(\nabla^2+m^2)\psi
\label{eq11}
\ee
\bb
\sqrt{M^2-m^2}(i{\gamma^\mu}\nabla_\mu-m)\psi=(\nabla^2+m^2)\varphi
\label{eq12}
\ee
\bb
\Box(A_\mu-{\pd_\mu}\Phi)=j^{el}_\mu
\label{eq13}
\ee
\bb
\Box{\pd^\mu}(A_\mu-{\pd_\mu}\Phi)=\Box{U}-{\pd^\mu}{j^\Phi_\mu}
\label{eq14}
\ee
\bb
\Box\Phi=U
\label{eq15}
\ee
where we have used the following notation:
\bb
j^{el}_\mu=-e\bar\varphi{\gamma_\mu}\varphi
\label{eq16}
\ee
\bba
j^\Phi_\mu&=&-{ie\over\sqrt{M^2-m^2}}(\bar\varphi{\nabla_\mu}\varphi+
\bar\psi{\nabla_\mu}\varphi-{\nabla^\star_\mu}\bar\varphi.\psi-
{\nabla^\star_\mu}\bar\psi.\varphi)- \nonumber \\ &-&
e\bar\psi{\gamma_\mu}\psi +{\pd^\nu}w_{\mu\nu}
\label{eq17}
\eea
Here $w_{\mu\nu}$ is an arbitrary antisymmetric tensor and the
last term in the righthand side of eq. (\ref{eq17}) expresses the
arbitrariness of the dependence of the current $j^\Phi_\mu$ from
its divergence (note that only ${\pd^\mu}j^\Phi_\mu$ enter the
field equations - eq. (\ref{eq14})).

As we mentioned above there are some auxiliary fields in our
model.
It is necessary to exclude them and obtain the final form of the
field equations. After simple calculations we have from
eqs. (\ref{eq11}) and (\ref{eq12}) :
\bb
(\nabla^2+m^2)(i{\gamma^\mu}\nabla_\mu+m)\varphi+
(M^2-m^2)(i{\gamma^\mu}D_\mu+m)\varphi=0
\label{eq18}
\ee
and from eqs. (\ref{eq14}), (\ref{eq15}) -
\bb
\Box{\pd^\mu}(A_\mu-{\pd_\mu}\Phi)=\Box^2\Phi-{\pd^\mu}j^\Phi_\mu
\label{eq19}
\ee
Furthermore from eqs. (\ref{eq13}), (\ref{eq19}) we can obtain
\bb
\Box^2\Phi={\pd^\mu}(j^{el}_\mu+j^\Phi_\mu)
\label{eq20}
\ee
It is easy to verify that the right hand side term in the last
equation is zero, i.e.,
\bb
{\pd^\mu}(j^{el}_\mu+j^\Phi_\mu)=0
\label{eq21}
\ee
as a result from field equations (\ref{eq11}) and (\ref{eq12}).
The calculations leading to this result one can see in the
Appendix. Then
\bb
\Box^2\Phi=0
\label{eq22}
\ee
and instead of eq. (\ref{eq19}) we have
\bb
\Box{\pd^\mu}(A_\mu-{\pd_\mu}\Phi)=-{\pd^\mu}j^\Phi_\mu
\label{eq23}
\ee
Now we can point out the final equations. These are the equations
(\ref{eq18}), (\ref{eq13}) and (\ref{eq22}) for the basic fields
and one of the eqs. (\ref{eq11}) or (\ref{eq12}) for the auxiliary
field $\psi$ (we ignore the field $U$). Certainly we must
consider our current $j^\Phi_\mu$ with excluded $\psi$. The rest
of the
equations (\ref{eq21}) and (\ref{eq23}) follow from the basic
ones.

To complete the description of our model, let us introduce new
basic fields using the above mentioned partial gauge
invariance. This we make with the help of the following
expressions:
\bb
\varphi(x)=\exp{ie\Phi(x)}.\phi(x);\ \ \ \ \ \ \
{A_\mu}(x)={\cal{A}_\mu}(x)+{\pd_\mu}\Phi
\label{eq24}
\ee
Then the operator $\nabla_\mu$ turns into $\pd_\mu$ in all
equations  and instead of (\ref{eq18}), (\ref{eq13}) and
(\ref{eq23}) we have:
\bb
{1\over{M^2-m^2}}(\Box+M^2)(i{\gamma^\mu}\pd_\mu+m)\phi+
e{\gamma^\mu}{\cal{A}_\mu}\phi=0
\label{eq25}
\ee
\bb
\Box{\cal{A}}_\mu=j^{el}_\mu
\label{eq26}
\ee
\bb
\Box\pd^\mu{\cal{A}}_\mu={\pd^\mu}j^{el}_\mu
\label{eq27}
\ee
respectively. As we can see we have excluded the current
(\ref{eq17}) from eq. (\ref{eq23}) using the identity (\ref{eq21}).
Then equation (\ref{eq27}) becomes a direct consequence of
eq. (\ref{eq26}).

\section{Quantization of the model}

We are going to quantize the model described above with the help
of the perturbation theory in Dirac (interaction) representation
\footnote {Sometimes this representation is called
Tomonaga-Shwinger representation.}
For this matter we will apply the approach proposed by Bogolubov
\cite{r8}. As is well known, this approach involves the
formulation of the quantization of the corresponding free field
theory. In our case we have from eq.(\ref{eq3}) the following
free spinor Lagrangean:
\bba
L^0(\phi)&=&{1\over\sqrt{M^2-m^2}}({\pd_\mu}\bar\phi{\pd^\mu}\psi+
{\pd_\mu}\bar\psi{\pd^\mu}\phi)+\bar\phi
(i/2\gamma^\mu \pd^{\!\!\!\! ^\leftrightarrow}_\mu
 + m)\phi+  \nonumber \\
 &+&
\bar\psi
(i/2\gamma^\mu \pd^{\!\!\!\! ^\leftrightarrow}_\mu
 -m)\psi-{m^2\over\sqrt{M^2-m^2}}(\bar\psi\phi+\bar\phi\psi)
 \label{eq28}
\eea

{\bf Remark}. Let us remind that when we are passing to the
quantum theory all products of the fields in the Lalgrangeans must
be understood as normal ones. This convention makes unnecessary
the use of any additional symbols for the normal product.

The Lagrangean for the field $\cal{A}_\mu$ is the well known
Lagrangean for the massless vector field and its quantization as
electromagnetic field in the Lorentz gauge is given, e.g., in
\cite{r8}.
The field $\Phi$ satisfying the eq. (\ref{eq22}) played an
important role in the conformal invariant QED and its quantum
theory is well known too (see for example the works \cite{r9}).
Moreover, we have seen that this field can be excluded from
our model with the help of the gauge transformations (\ref{eq24}).
On the other hand let us remind that the field
$\psi(x)$ is an auxiliary spinor field which has to be excluded
too. However, the Lagrangean (\ref{eq28}) is more suitable for
canonical quantization (containing first derivatives
only)
than one from (\ref{eq6}).
That is why here we will consider the theory with the Lagrangean
(\ref{eq28}) only. The free field equations in our case have the
form:
\bb
{1\over\sqrt{M^2-m^2}}(i\gamma^\mu\pd_\mu+m)\phi=
(\Box+m^2)\psi
\label{eq29}
\ee
\bb
{1\over\sqrt{M^2-m^2}}(i\gamma^\mu\pd_\mu-m)\psi=
(\Box+m^2)\phi
\label{eq30}
\ee
The corresponding canonical momenta are
\bb
\pi_\psi(x)={1\over\sqrt{M^2-m^2}}\pd_0\bar\phi(x)+{i\over2}
\psi^\star(x)=(\pi_{\bar\psi}(x))^\star\gamma^0
\label{eq31}
\ee
\bb
\pi_\phi(x)={1\over\sqrt{M^2-m^2}}\pd_0\bar\psi(x)+{i\over2}
\phi^\star(x)=(\pi_{\bar\phi}(x))^\star\gamma^0
\label{eq32}
\ee
The anticommutator of the field $\phi(x)$ we denote as
\bb
\Gamma_{\alpha\beta}(x)=\{\phi_\alpha(x),\bar\phi_\beta(0)\}
\label{eq33}
\ee
($\{...\}$ means anticommutator).
From equations (\ref{eq29}) and (\ref{eq30}) we have the
following equation for the function $\Gamma_{\alpha\beta}(x)$:
\bb
(\Box+M^2)(i\gamma^\mu\pd_\mu+m)\Gamma(x)=0
\label{eq34}
\ee
Furthermore using the canonical commutation relations:
\bb
\{\pi_\phi(x),\phi(y)\}_{x_0=y_0}=\{\pi_\psi(x),\psi(y)\}_{x_0=y_0}=
i\delta^3(x)\delta_{\alpha\beta}
\label{eq35}
\ee
we have the following initial conditions for the function $\Gamma$
\bba
&\Gamma_{\alpha\beta}(x)\vert_{x_0=0}=\pd_0\Gamma_{\alpha\beta}(x)
\vert_{x_0=0}=0  \nonumber \\
&\pd_0^2\Gamma_{\alpha\beta}(x)\vert_{x_0=0}=(M^2-m^2)\gamma^0_{\alpha
\beta}\delta^3(x)
\label{eq36}
\eea
To obtain the latter we have used the anticommutation relations
between the fields $\psi$ and $\phi$, and the equations they
satisfy. These calculations are very simple and we omit them here.

Now it is easy to obtain our function $\Gamma$:
\bb
\Gamma_{\alpha\beta}(x)=-i(i\gamma^\mu\pd_\mu-m)_{\alpha\beta}
[D_m(x)-D_M(x)]
\label{eq37}
\ee
where $D_m$ and $D_M$ are the scalar Pauli-Jordan functions
with masses $m$ and $M$ respectively. Analogously one can obtain
the rest of the singular functions from which we will write down
here the
causal Green function only:
\bb
\Gamma^c_{\alpha\beta}(x)=-i(i\gamma^\mu\pd_\mu-m)_{\alpha\beta}
[D^c_m(x)-D^c_M(x)]
\label{eq38}
\ee
$D^c_m$ and $D^c_M$ are corresponding scalar causal Green
functions for the masses $m$ and $M$:
$$D^c_m(x)={1\over{(2\pi)^4}}\int{e^{-ikx}\over{m^2-k^2-i\varepsilon
}}d^4k$$
Our causal Green function (\ref{eq38}) is normalised in such a way
that it coincides with the propagation function:
\bb
<T\phi_\alpha(x),\bar\phi_\beta(0)>_0=\Gamma^c_{\alpha\beta}(x)
\label{eq39}
\ee
The formulae (\ref{eq37}) and (\ref{eq38}) show us that the free
quantum theory of our field $\phi$ is containing indefinite
metric in state space. This fact becomes more clear if we
consider the concrete form of the general solution for the field
$\phi$ . It has the form
\newpage
\bba
\phi_\alpha(x)&=&\phi^0_\alpha(x)+ \nonumber \\
&+&{1\over{(2\pi)^{3/2}}}\sum_{{\it{r}}=1,2}
\int\{[\sqrt{\Omega(M+m)\over2M}a^{+{\it{r}}}({\bf{p}})
v^{+{\it{r}}}_\alpha(\Omega,{\bf{p}})+ \nonumber \\
&+&
\sqrt{\Omega(M-m)\over2M}
c^{-{\it{r}}}({\bf{p}})
v^{-{\it{r}}}_\alpha(\Omega,{\bf{p}})]e^{ipx}+ \nonumber \\
&+&[\sqrt{\Omega(M-m)\over2M}b^{+{\it{r}}}({\bf{p}})
v^{+{\it{r}}}_\alpha(\Omega,
{\bf{p}})+  \nonumber \\
&+&
\sqrt{\Omega(M+m)\over2M}d^{-{\it{r}}}({\bf{p}})v^{-{\it{r}}}_\alpha
(\Omega,{\bf{p}})]e^{-ipx}\}{d^3p\over\Omega} \label{eq40}  \\
\nonumber\\
&&\Omega=\sqrt{{\bf{p}}^2+M^2} \nonumber \\
&&{\bf{p}}=(p_1,p_2,p_3) \ \ \ \ \ \ \ {\bf{p}}^2=\sqrt{p_1^2+
p_2^2+p_3^2}  \nonumber
\eea
Here $\phi^0_\alpha(x)$ is the usual free spinor field satisfying
Dirac equation
$$(i\gamma^\mu\pd_\mu+m)\phi^0(x)=0$$
and the following anticommutation relation
\bb\{\phi^0_\alpha(x),\bar\phi^0_\beta(y)\}=
-i(i\gamma^\mu\pd_\mu-m)_{\alpha\beta}D_m(x-y)
\label{eq41}
\ee
The mode-vectors $v^{\pm\alpha}(\Omega,{\bf{p}})$ fulfill the
usual spinor identities which in our case have the form
\bb
(\gamma^\mu p_\mu \mp M)v^{\pm{it{r}}}(\Omega,{\bf{p}})=0
\label{eq42}
\ee
Moreover this mode-vectors satisfy the normalization condition
which can be written down in the following two forms:
\bba
&&\sum_{\alpha=1}^4(v^{\pm{\it r}}_\alpha(\Omega,{\bf p}))^\star
v^{\pm{\it s}}_\alpha(\Omega,{\bf p})=\delta_{\it rs}
\label{eq43} \\
&&or  \nonumber \\
&&\sum_{\alpha=1}^4(\overline{v^{\pm{\it r}}_\alpha(\Omega,{\bf p})})
v^{\pm\it s}_\alpha(\Omega,{\bf p})=
\pm{M\over\Omega}\delta_{\it rs}
\label{eq44}
\eea
The relation
\bb
\sum_{\alpha=1}^4(v^{\pm{\it r}}_\alpha(\Omega,\pm{\bf p}))^\star
v^{\mp{\it s}}_\alpha(\Omega,\mp{\bf p})=0
\label{eq45}
\ee
is a condition for the orthogonality of the mode-vectors and
finally the relation
\bb
\sum_{{\it r}=1,2}v^{\pm{\it r}}(\Omega,{\bf p})
 (\overline{v^{\pm{\it r}}(\Omega,{\bf p})})=
{{\gamma^\mu p_\mu\pm M}\over{2\Omega}}
\label{eq46}
\ee
expresses the completeness of these vectors.

The quantities $a^{+{\it r}}({\bf p})$, $b^{+{\it r}}({\bf p})$,
$c^{-{\it r}}({\bf p})$ and $d^{-{\it r}}({\bf p})$
entering the general solution (\ref{eq40}) are operators in the
quantum theory and they must satisfy the following
anticommutation relations:
\bba
\{a^{+{\it r}}({\bf p}),a^{-{\it s}}({\bf q})\}&=&
-\Omega\delta_{\it rs}\delta^3({\bf p-q}) \nonumber \\
\{b^{+{\it r}}({\bf p}),b^{-{\it s}}({\bf q})\}&=&
-\Omega\delta_{\it rs}\delta^3({\bf p-q}) \label{eq47}  \\
\{c^{+{\it r}}({\bf p}),c^{-{\it s}}({\bf q})\}&=&
-\Omega\delta_{\it rs}\delta^3({\bf p-q}) \nonumber \\
\{d^{+{\it r}}({\bf p}),d^{-{\it s}}({\bf q})\}&=&
-\Omega\delta_{\it rs}\delta^3({\bf p-q}) \nonumber
\eea
(all other anticommutators are zero) according to the
(\ref{eq33}). In the above written formulas we have
made the following notation
\bba
&a^{-{\it r}}({\bf p})=(a^{+{\it r}}({\bf p}))^\star \nonumber \\
&b^{-{\it r}}({\bf p})=(b^{+{\it r}}({\bf p}))^\star \label{eq48} \\
&c^{+{\it r}}({\bf p})=(c^{-{\it r}}({\bf p}))^\star \nonumber \\
&d^{+{\it r}}({\bf p})=(d^{-{\it r}}({\bf p}))^\star \nonumber
\eea
The new quantities $(a^{+{\it r}}({\bf p}))^\star, (b^{+{\it
r}}({\bf p}))^\star, (c^{-{\it r}}({\bf p}))^\star, (d^{-{\it
r}}({\bf p}))^\star $ appear in the field $\bar\phi_\alpha(x)$
Dirac conjugated to the field (\ref{eq40}) and the sign $\star$
has
the meaning of Hermitean conjugation.

Now let us describe the structure of our state space $S$. First
of all $S$ is containing the subspace ${\it H}$ which coincides
with the Fock space of the free quantum field $\phi^0(x)$. This
is the usual state space of the free quantum spinor field with
positive metric. Its vacuum state $\mid 0>$ is common to the
whole space $S$ and satisfies the following additional conditions:
\bba
a^{-{\it r}}({\bf p})\mid 0>&=&b^{-{\it r}}({\bf p})\mid 0>=
c^{-{\it r}}({\bf p})\mid 0>=d^{-{\it r}}({\bf p})\mid 0>=0
\label{eq49} \\
<0\mid a^{+{\it r}}({\bf p})&=&<0\mid b^{+{\it r}}({\bf p})=
<0\mid c^{+{\it r}}({\bf p})=<0\mid d^{+{\it r}}({\bf p})=0
\nonumber
\eea
Then we can see that $S$ contains an additional subspace ${\it
G}$
in which the basis is formed from all monomials of creation
operators
($a^{+{\it r}}({\bf p}), b^{+{\it r}}({\bf p}), c^{+{\it
r}}({\bf p}),\\ d^{+{\it r}}({\bf p})$) acting on the
vacuum. The corresponding operators with the sign "-" are the
annihilation operators. According to the our commutation relations
given above, we can do the following conclusions:

i) The metric of the space ${\it G}$ given with the usual scalar
product of the Fock space is indefinite.

ii) The spaces ${\it H}$ and ${\it G}$ are orthogonal to each other
and
$$S={\it H}\otimes{\it G}$$
It is obvious that the vectors of the space $S$ can not be used as
physical states because of the presence of the ghosts from
subspace ${\it G}$.

However, we can define the physical space to coincide with the
subspace ${\it H}$ and consider only its vectors as edge physical
states. Then it is easy to verify that
$$<ph'\mid\bar\phi(x)\gamma^\mu\phi(x)\mid ph>=
<ph'\mid\bar\phi^0(x)\gamma^\mu\phi^0(x)\mid ph>$$
according to the eq.(\ref{eq40}) and of course
\bb
\pd_\mu<ph'\mid\bar\phi(x)\gamma^\mu\phi(x)\mid ph>=0
\label{eq50}
\ee
where $\mid ph>$ and $\mid ph'>$ are two arbitrary states from the
physical space ${\it H}$. Moreover, in this case the relation
(\ref{eq50}) define the space ${\it H}$ and can be considered as
a defining condition for the physical states.

Passing to our model with interaction it is naturally to
generalize the condition (\ref{eq50}) for the definition of the
physical states in the interaction case. This we can do defining
the latter with the help of the relation, formally coinciding with
relation (\ref{eq50}) but with $\phi(x)$ satisfying the equations
(\ref{eq25})-(\ref{eq27}). Such definition is in accordance, except
for the free case, with the quantum theory in the interaction
representation, where the spinor physical space coincides with the
space ${\it H}$.

As we mentioned above we consider our model only on the physical
space. This means that all matrix elements of operators and their
products having some physical meaning must by taken between
physical states only. Then the
nonphysical states such as ghost ones will give contribution to
the intermediate virtual states only.

Now we are going to see what happens with our model in the
physical space. According to the condition (\ref{eq50}) we can
obtain there the form of the equations (\ref{eq26}) and
(\ref{eq27}):
\bb
<ph'\mid\Box{\cal A}_\mu(x)-j^{el}_\mu(x)\mid ph>=0
\label{eq51}
\ee
and
\bb
<ph'\mid\Box\pd^\mu{\cal A}_\mu(x)\mid ph>=0
\label{eq52}
\ee
The last equation shows  us that in  the chosen physical space
automatically appears the gauge condition such as
$$\Box\pd^\mu{\cal A}_\mu =0$$
This condition contains the Lorentz gauge fixing $$\pd^\mu{\cal
A}_\mu=0$$
which means that the our physical space has a subspace in which
equation (\ref{eq51}) coincides with the Maxwell equation in the
Lorentz
gauge. The vectors belonging to this restricted physical space
are denoted here as $\mid ph_0>$. Then instead of
eqs. (\ref{eq51}) and (\ref{eq52}) we have
\bba
&&<ph'_0\mid\Box{\cal A}_\mu(x)-j^{el}_\mu(x)\mid ph_0>\equiv
\nonumber \\
&&\equiv  <ph'_0\mid\pd^\nu{\it F}_{\nu\mu}(x)-j^{el}_\mu(x)\mid
ph_0>=0  \label{eq53}  \\
&& F_{\mu\nu}=\pd_\mu{\cal A}_\nu - \pd_\nu{\cal A}_\mu
\nonumber
\eea
and
\bb
\pd^\mu<ph'_0\mid {\cal A}_\mu(x)\mid ph_0>=0
\label{eq54}
\ee

The construction described here obtains its concrete form in the
interaction representation, where, as is well known, the
Lagrangean and the other physical quantities are expressed
through the
corresponding free quantum fields in the Heisenberg
representation. In this case the mentioned above restricted
physical space ${\it H}_0$ in which equations (\ref{eq53}) and
(\ref{eq54}) take place, coincides with the following space with
nonnegative metric:
\bb
{\it H_0=H\otimes R_a}
\label{eq55}
\ee
where the space ${\it R_a}$ can be defined as
\bba
&&{\it R_a}\equiv\bigoplus_{N\geq 0}\bigotimes_{{\it k}=0}^N
{\bf l}^1_{\it k} \label{eq56} \\
&&{\bf l}^1_0= \mid l^1_0> \equiv \mid 0> \nonumber
\eea
and where ${\bf l}^1_{\it k}$ are the one particle spaces for the
field
${\cal A}_\mu(x)$. The arbitrary one particle state in this case
has the form
\bb
\mid l^1>=\int{\cal A}^+_\mu(x)l^\mu(x)d^4x\mid 0>
\label{eq57}
\ee
where ${\cal A}^+_\mu(x)$ is the positive frequency part of the
field ${\cal A}_\mu(x)$ and $l^\mu(x)$ are arbitrary test
functions fulfilling the condition:
\bb
\pd_\mu l^\mu(x)=0
\label{eq58}
\ee
i.e., $$\pd^\mu{\cal A}^-_\mu(x)\mid l'>={1\over i}\int\pd^\mu
D^-_0(x-y)l_\mu(y)
d^4x \equiv 0$$
Here ${\cal A}^-_\mu(x)$ and $D^-_0(x-y)$ are the negative
frequency
parts of the field ${\cal A}_\mu(x)$ and the massless Pauli
Jordan
function respectively.

Our physical space is the subspace of the whole quantum state
space
$Q$ which in the considered representation has the structure
\bb
Q=S\otimes F_A ; \ \ \ \ \ {\it H}_0\subset Q
\label{eq59}
\ee
where $F_A$ is the Fock space of the quantum vector field ${\cal
A}_\mu(x)$
satisfying the d'Alambert equation.

Now we can formulate the main result of the present paper:

The quantum model with the action (\ref{eq2})
and Lagrangeans (\ref{eq3}), (\ref{eq4}) and (\ref{eq10}), leading
to the field equations with the higher derivatives (\ref{eq25}),
(\ref{eq26}) and (\ref{eq27}), contains in its own state space $Q$
the subspace ${\it H}_0$ with nonnegative metric in which our
model coincides with the spinor QED. From this point of view the
latter is immersed in our model described above. It is necessary
to mention that this QED is in regularized form, because the
spinor propagation function is given by expression (\ref{eq38})
and, as is easy
to see, it coincides with the Pauli-Villars
regularized one. That is why the Feynman diagrams
are free from divergences in our case. In the next section we are
going to discuss these questions.

\section{Regularized QED}

In this section we would like to compare the described above
regularized QED (RQED) which is immersed in our model, with the
standard
one. This comparison we are going to do on the level of Feynman
diagrams and it helps to understand better our reasons to
consider such theory.

As we have seen, the physical space of RQED in the interaction
representation is composed from Fock spaces of the free spinor
$\phi^0_\alpha(x)$ and free photon ${\it A}_\mu(x)$ fields in
Lorentz gauge. This space is with nonnegative metric and gives us
the possibility to do the next steps to obtain the physical space
of the
free electrons, positrons and photons, e.g., passage to the Coulomb
gauge fixing, factorization of the zero norm state vectors and so
on (see ref. \cite {r9}); the steps usually done in
standard QED too.
Then it is easy to notice, that our physical space is the same as
in standard QED. There is no difference also between the
photon
propagation functions in the two theories. In our case this
function is $$\Delta^c_{\mu\nu}(x)=\eta_{\mu\nu}D^c_0(x)$$
where
$\eta_{\mu\nu}$ is the Minkowski metric tensor
and
$D^c_0(x)$ is the causal Green function of the d'Alambert equation.

{\bf Remark}. In our model we have used the Lorentz gauge fixing
as the most simple. However, there are no obstacles to use
other forms of the photon part of our RQED. For instance if we had
used, instead of Lagrangean (\ref{eq4}), a new one as follows:
$$L_v(A)=-1/2\pd^\mu(A_\nu-\pd_\nu\Phi)\pd_\mu(A^\nu-\pd^\nu\Phi)
+{\kappa\over2}\pd^\mu(A_\nu-\pd_\nu\Phi)\pd^\nu(A_\mu-\pd_\mu\Phi)$$
the corresponding equations for the photons on the physical states
would have obtained
the form $$\Box{\cal A}_\nu-\kappa\pd_\nu\pd^\mu{\cal A}_\mu =
j^{el}_\nu$$ with the following gauge fixing condition
$$\Box\pd^\mu{\cal A}_\mu=0$$ This form of the equation for the
electromagnetic field ${\cal A}_\mu$ is well known too (see ref.
\cite {r10})

The main difference between RQED and the standard QED is in  the
form
of the spinor propagation function. As it is seen from
eq. (\ref{eq38}) the form of
our spinor propagation function coincides with
the well known
Dirac one but regularized with the help of the Pauli-Villars
regularization without performing the limit $M\rightarrow
\infty$.
In the momentum representation the function (\ref{eq38}) has the
behaviour $p^{-3}$ for $p\rightarrow\infty$, so RQED is free
of ultraviolet divergences. However, a new parameter $M$ appears
here. For the understanding of the influence of $M$ over
the RQED we are going to consider all Feynman diagrams arranged in
two groups. In the first group we put all diagrams which remain
finite in the limit $M\rightarrow\infty$. These diagrams
correspond to converging ones in the standard QED. Here it
is possible to choose $M$ large enough so that the corresponding
diagrams from the two QED's will coincide with each other on such
a range of the values of the external momenta, which could have
any experimental significance. That is why we can say that
the first group of diagrams RQED differs from
standard QED only for very big external momenta, the range of
values of which is defined by the value of the mass parameter
$M$.

In the second group of Feynman diagrams we put those which
correspond to the divergent diagrams of QED.
These diagrams increase unlimitedly for large $M$.
In the standard QED over these
diagrams one applies the renormalization procedure. In the our
case, because of the finiteness there is no similar necessity
in the RQED. That is why  all nonuniqueness appearing in QED
after the infinite renormalization
is described here through the
parameter $M$ only. In our opinion this is the main difference
between the two considered here QED's. We would like to
demonstrate this difference with the help of some example. For
this we choose the second order electron-positron self-energy
Feynman diagram, i.e., one of the basic divergent diagrams in
standard QED.

As usual the mentioned diagram we denote as
${\Sigma}_{\alpha\beta}(p)$, where $p$ is the external
electron-positron momentum. Without giving here the calculation
which is well known, we will start from the following expression
for this quantity in our case:
\bb
{\Sigma}_{\alpha\beta}(p)={e^2\over{8\pi^2}}\int_{0}^{1}d\xi
(2m-\gamma^\mu p_\mu \xi)_{\alpha\beta}ln{{\xi p^2-M^2+i\epsilon}
\over{\xi p^2-m^2+i\epsilon}}
\label{eq60}
\ee
The integration in the right hand side of the last equation can be
taken and for us is interesting the result when $M^2\gg p^2$,
i.e., in the low energy range. Then we have
\bb
{\Sigma}(p)={e^2\over{16\pi^2}}(4m-\gamma^\mu p_\mu)ln{M^2\over
m^2} +\Sigma'(p)
\label{eq61}
\ee
where we have denoted by $\Sigma'(p)$ the part which does not
increase with $M$
\bb
\Sigma'(p)={e^2\over{8\pi^2}}\int_{0}^{1} d\xi(2m-\gamma^\mu
p_\mu) ln{m^2\over{m^2-\xi p^2}} + O({1\over M})
\label{eq62}
\ee
The corresponding diagram in the standard QED after
renormalization has the form
\bb
{\Sigma}(p)=c_1 (\gamma^\mu p_\mu-m) +c_2+\Sigma'(p)
\label{eq63}
\ee
where $\Sigma'(p)$ is the same as in eq. (\ref{eq62}) (see ref.
\cite{r8}). Here $c_1$ and $c_2$ are arbitrary finite constants in
the result of the renormalization. Comparing eq. (\ref{eq63}) with
eq. (\ref{eq61}) we can see that in our case both constants
depend on $M$ as follows
$$c_1={{-e^2}\over{16\pi^2}}ln{M^2\over m^2} \ \ \ \
c_2={{3e^2m}\over{16\pi^2}}ln{M^2\over m^2}$$
This means that in RQED we have no nonuniqueness appearing in
the process of the calculation of such a diagram.

{\bf Remark}: It is possible to do finite renormalization of our
model. Then the analogous nonuniqueness will appear in RQED too.
However, the finiteness of the considered diagrams makes such
procedure unnecessary.

The only quantities which can be considered as arbitrary
parameters are the mass parameters $M$ and $m$. Recalling that
${\Sigma}_{\alpha\beta}(p)$ is the second order correction to the
free spinor operator we can write down the corresponding
corrected
Green function in the momentum representation:
\bb
G(p)={1\over{{1\over{M^2-m^2}}(p^2-M^2)(\gamma^\mu
p_\mu-m)-\Sigma (p)}}
\label{eq64}
\ee
Now we can calculate the pole-point of this function which will
represent the electron-positron physical mass $m_{ph}$. Up to
second order, this mass can be obtained in the form:
\bba
m_{ph}&=&m+\delta m  \nonumber \\
\delta m &=& -{{3e^2m}\over{16\pi^2}}ln{M^2\over m^2}-{5e^2m\over 32\pi^2}
\label{eq65}
\eea

It is well known that the point $p^2=m^2$ is a branching point of
$\Sigma'(p)$ and because of the absence of the renormalization it
is impossible to set $m_{ph}=m$. That is why there are bare $m$
and dressed $m_{ph}$ electron-positron masses in RQED, connected by
expression (\ref{eq65}). The dressed mass $m_{ph}$ is a pole of the
Green function (\ref{eq64}) and it is less than the bare mass $m$
which is a branching-point of the same Green function. Here these
points are automatically different and there is no necessity in
RQED to introduce small photon mass to reach this difference, as
it takes place in the standard QED. Perhaps this is the most
significant property. If we reformulate it, this property means
that there are no infrared divergences in the considered diagram
(see for this ref.\cite{r8}).

\section {Conclusion}

We defined a model with higher derivatives. Our analysis
shows that in the suitable chosen physical subspace of states,
this model coincides with QED. The only difference between QED and
our model (calling here RQED) is the usage of the regularized
causal spinor Green function in the latter. However, in our theory
we can not take $M\rightarrow\infty$, because the mass parameter
$M$ enter the initial spinor part of the Lagrangean and it has a
finite value. After simple analysis we saw that RQED is free from
ultraviolet divergences. Choosing the parameter $M$ sufficiently
large, we can say that RQED and QED have the same behaviour in
the low energy range. We are hopeful that RQED will have a
behaviour in the high energy range better than QED, because of
the faster decrease of our spinor propagator for large
momentum.

On the example of the spinor selfenergy second order Feynman
diagram we saw that there is no infrared divergence too.

From the obtained results we can see that there exists an
alternative way to consider given particle theory in which the
field operator describes not only the physical one. Then the
physical theory turns out as immersed into wider theory, such
that the latter has contribution to the intermediate (virtual)
states only. We have demonstrated here this approach on the well
known theory for comparison. However, there are theories such
as nonrenormalizable ones, where this approach could turn
out to be the only possible. For example, applying our propagator
to the four-fermions interaction, the corresponding theory will
have no divergences at all.

Finally the author would like to thank M. Stoilov for useful
discussion and V. Dobrev for reading the manuscript.

\bigskip\bigskip

\centerline{\bf APPENDIX}

First of all let us write down the full current $j^{el}_\mu+j^\Phi_\mu$
in the terms of fields defined in eq. (\ref{eq24}). Then we have
$$j^{el}_\mu+j^\Phi_\mu=-e{\pd^\mu}\bar\phi{\gamma_\mu}\phi-
e{\pd^\mu}\bar\xi{\gamma_\mu}\xi-$$
$$-{ie\over\sqrt{M^2-m^2}}(\bar\phi{\pd_\mu}\xi+\bar\xi{\pd_\mu}\phi
-{\pd_\mu}\bar\phi.\xi-{\pd_\mu}\bar\xi.\phi)$$
where by $\xi(x)$ we have denoted
$$\xi(x)=\exp[{-ie\Phi(x)}].\psi(x)$$
Then the equations (\ref{eq11}) and (\ref{eq12}) can be written
down as follows:
$$\sqrt{M^2-m^2}(i{\gamma^\mu}\pd_\mu+m)\phi=(\Box+m^2)\xi-
e{\cal A}_\mu{\gamma^\mu}\phi \eqno(A.1)$$
$$\sqrt{M^2-m^2}(i{\gamma^\mu}\pd_\mu-m)\xi=(\Box+m^2)\phi
\eqno(A.2) $$
We can do the verification of eq. (\ref{eq21}) with direct
calculations. For this let us write down the Dirac conjugates to
eqs. (A.1) and (A.2)
$$\sqrt{M^2-m^2}(-i{\pd_\mu}\bar\phi{\gamma^\mu}+m\bar\phi)=
(\Box+m^2)\bar\xi-e\bar\phi{\gamma^\mu}{\cal A}_\mu
\eqno (A.3)$$
$$\sqrt{M^2-m^2}(-i{\pd_\mu}\bar\xi{\gamma^\mu}-m\bar\xi)=
(\Box+m^2)\bar\phi \eqno(A.4)$$
Now we must multiply eqs. (A.1) and (A.3) by $\bar\phi$ from
left and $-\phi$ from right respectively and then add the results.
Analogously we must proceed with eqs. (A.2) and (A.4). It
is easy to see then, that the sum of the two obtained in this
manner identities coincides with the expression:
$${\pd^\mu}(j^{el}_\mu+j^\Phi_\mu)=0$$


\bigskip\bigskip

\newpage


\begin{thebibliography}{99}
\bibitem{r1}T Nakamura  and  S Hamamoto, "Higher  Derivatives  and
Canonical Formalisms" hep-th/9511219, (1995) and references therein.
\bibitem{r2} M Ostrogradski, Mem. Ac. St. Petersbourg, {\bf 14}
,(1850) 385
\bibitem{r3}L V Belvedere,  D L P Amaral,   N A Lemos   "Canonical
Transformations  in  a  Higher-Derivative  Field  Theory",
hep-th/9512002 (1995).
\bibitem{r4}C G Bollini, L E Oxman, M Rocca, J. Math. Phys., {\bf 35},
(1994) 4429 and references therein.
\bibitem{r5}T Hamazaki and T Kugo, Progress  Theor.  Phys., {\bf 92},
(1994) 645
\bibitem{r6}D Z Freedman, K Johnson and J I Latorre,  Nucl.
Phys., {\bf B371}, (1992) 329
\bibitem{r7}S W Hawking  "Who's Afraid  of   (Higher   Derivatives)
Ghosts?". In "Quantum Field Theory and Quantum Statistics", vol.2
p.129,  ed. Batalin I A et al. (Cambridge, 1985)
\bibitem{r8}N N Bogolubov and D V Shirkov, {\it Introduction in
Quantum Field Theory} (Moscow, {\it Nauka}, 1973)
\bibitem{r9} D.Zwanziger, Phys.Rev., {\bf D17}, (1978) 457 ;G
Sotkov, D Stoyanov and S Zlatev, Commun. JINR P2-128000 (1979);
G M Sotkov, D T Stoyanov, J. Phys. A: Math. Gen. {\bf 16} (1983)
2826
\bibitem{r10}C Itzykson and G-B Zuber "Quantum Field Theory"
McGraw-Hill Book Company (New York, 1978)
\end{thebibliography}
\end{document}